# Pseudoscopic imaging in a double diffraction process with a slit

**José J. Lunazzi and Noemí I. Rivera**

*Universidade Estadual de Campinas, Institute of Physics,C.P.6165/13084-100 Campinas-SP Brazil*

*lunazzi@ifi.unicamp.br*

*http://www.ifi.unicamp.br/~lunazzi*

**Abstract:** Pseudoscopic images that keep the continuous parallax are shown to be possible due to a double diffraction process intermediated by a slit. One diffraction grating acts as a wavelength encoder of views while a second diffraction grating decodes the projected image. The process results in the enlargement of the image under common white light illumination.
©2002 Optical Society of America

___________________________________________________________________________________________

___________________________________________________________________________________________

## 1. Introduction

Refractive or reflective optics cannot render a large parallax field due to their limited aperture. Direct pseudoscopic images are uncommon [1]. After the development of holographic images it was possible to appreciate the benefits of having images that may render a wide field of view while keeping the continuous parallax, allowing the observer to "look around" the scene to obtain the maximum of its visual information.  Holography and diffractive imaging may render continuous parallax exclusively under monochromatic light or through some process that renders the final image monochromatic at least over the horizontal field of view.  We demonstrate in this paper that diffraction can be combined with a simple imaging process to obtain images for binocular viewing whose continuous horizontal parallax is due exclusively to diffractive elements. We demonstrated in two previous papers [2,3] that the ability of wavelength-encoding a continuous sequence of views may easily be obtained by simple diffraction at a grating and stated that it may also be decoded at a double diffraction process [4,5] intermediated by a lens and a slit. We employed a lens at the symmetry center in [4] in order to get more luminosity and sharpness, at the expense of a more complicated ray-tracing problem. In this paper we demonstrate now how a second diffraction grating symmetrically located in respect to a simple slit is the natural way of decoding the light distribution previously coded in a first diffraction process. Symmetry properties are enough to demonstrate the generation of a pseudoscopic image, a kind of image that was only known from stereo photographic or holographic processes but not known in diffractive optics.

**2. Ray-tracing in a double diffraction imaging process**

Our system consists of two identical diffraction gratings $DG1$, $DG2$, symmetrically located at both sides of an aperture $a$ (see Fig.1).

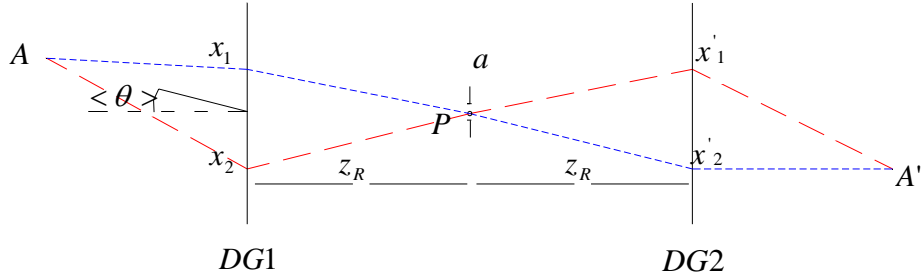

Fig. 1. ray-tracing for the symmetrical image of a point white-light object.

The plane of the figure corresponds to a horizontal plane containing the center $P$ of the aperture, which we considered as coordinate center, while the lines of the grating are in a vertical direction. An object of white or gray tonality is illuminated by common white light diffusing at a very wide angle, such as, for example, from point $A$ whose coordinates are $x, y, z$. We consider the part of the beam reaching the grating, which, after diffraction, travels toward the aperture. $Z_R$ is the distance from the slit to the first grating, made equal to the distance from the slit to the second grating, a symmetry condition. If, for example, light reaches the grating at the point where it intersects the perpendicular to the aperture, it means that the mean direction $<\theta>$ of the object light must satisfy:

$$sin <\theta> = <\lambda> .\nu \qquad (1)$$

$\nu$ being the inverse of the grating period and $<\lambda>$ the mean value of the wavelength of the visible spectrum, about 0.55 µm. A rather different situation could have been chosen, where light impinges the grating normally, which could be analyzed with minor changes. The aperture is a simple vertical slit and we consider that diffraction by the aperture can be neglected because it does not affect resolution as much as the extension of the slit would. So we can call the process a double diffraction process or, more properly, a diffraction-absorption-diffraction process. Light rays from an object point $A$ from which light is diffused in all directions may reach, for example, two different points $x_1$, $x_2$ at the first grating. Only one wavelength value allows the light to travel from the grating to the point $P$ at the center of the aperture, satisfying the grating equation. We have then for each point at the grating:

$$sin\theta_i - sin\theta_d = \lambda.\nu \qquad (2)$$

where $\theta_i$ represents the angle of incidence of light traveling from point $A$ to points on the grating, $\theta_d$ represents the angle of diffraction for light that travels from points on the grating to point $P$. $\lambda$ represents the wavelength value corresponding to each ray. It must be noticed that in this situation the center of the aperture receives only a single ray that corresponds to a specific wavelength value. Due to the symmetry of the optical elements all rays reach the second grating at points such as $x'_1$, $x'_2$ that are symmetric to the points from where they left the first grating. When the rays reach the second grating, only one of the two first diffraction orders is considered, the one that allows to keep the central symmetry through point $P$. It creates a situation of perfect symmetry that renders an image point $A'$. We can see this through the same Eq.(2) by using the appropriate new corresponding angles. The case is not symmetric for the other order, which can nevertheless bring an orthoscopic image, a case that will be described elsewhere. Geometrically describing the diffraction direction of a given wavelength we obtain the relationship which describes the light path:

$$(x-x_1)/\sqrt{(x-x_1)^2 + z^2} - x_1/\sqrt{x_1^2 + z_R^2} = \lambda\nu \qquad (3)$$

**3. The pseudoscopic imaging case**

The symmetry that we described demonstrates that a pseudoscopic real image may be obtained which is symmetric to the object, considering the point of symmetry being the center of the aperture. We show this situation for two object points $A$ and $B$ located at different depth positions. See Fig.2.

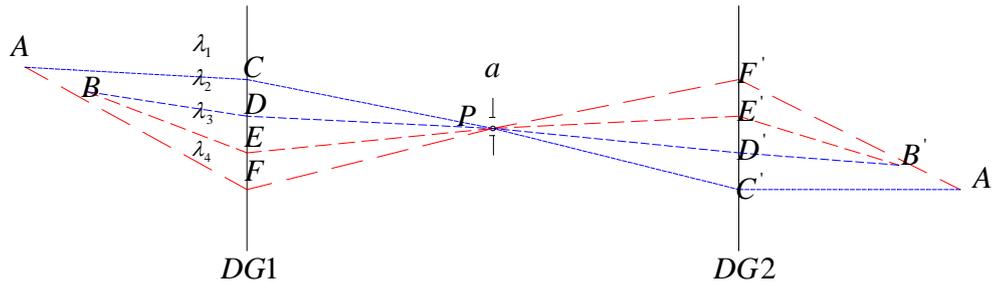

Fig. 2. ray-tracing scheme for the depth inverted image.

Four rays impinge on the first grating at points $C$, $D$, $E$, $F$, whose wavelengths $\lambda_i$ have subscript numbers chosen in such a way that, if one is greater than another, it means that the indicated wavelengths are correspondingly greater than one another. Points $C'$, $D'$, $E'$, $F'$, where light reaches the second grating, are clearly symmetric to the corresponding points where light reached the first grating. All wavelengths are recombined at image points according to the spectral sequence, spreading from there in an inverted sequence. The observer will see the image point $A'$ as being closer to him than image point $B'$, therefore receiving a depth-inverted view of the object. The image resembles the previous cases of pseudoscopic images obtained with holographic screens [6] [4].

**4. Spectral distribution and parallax effects on the image**

When an observer is included on the ray tracing, his viewpoint breaks the symmetry of the ray scheme and the first consequence is the false coloring of the scene. The scheme of Fig.3 explains that the observer sees the scene in a horizontal sequence of colors when looking around the image.

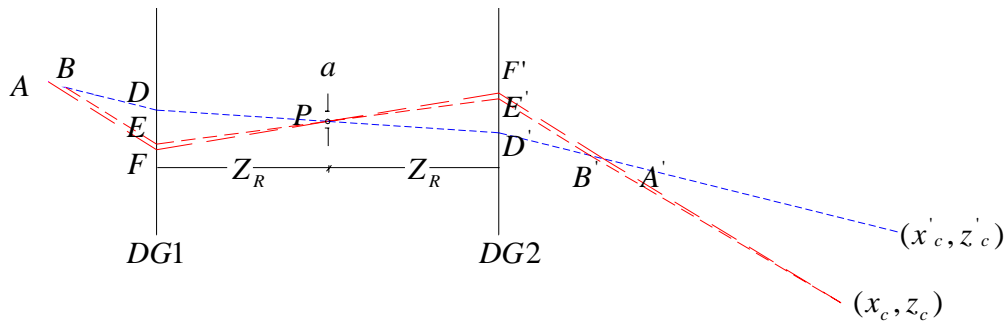

Fig. 3. Ray-tracing scheme for rays reaching an observer. Two object points A,B are represented.

When his observing point passes from position $(x_c, z_c)$ to $(x'_c, z'_c)$, for example, he receives a different wavelength. The restrictive condition of eq.3 makes the light distribution coming from the center of the slit to follow the spectral sequence. As expressed before, the extreme wavelength values determine the angle of viewing, and any ray that converges to an imaging point and is directed to the eye must obey the spectral sequence relationship. The reduced extension of the pupil makes light at the eye to have very limited bandwidth. All wavelengths recombine at the image points and the original spectral distribution of the object is reconstructed there, but, since the observation is made from a certain distance, the observer can only see one pure spectral color at a time. The viewing of two object points also corresponds to two wavelengths (see Fig.3) for the observer at $(x_c, z_c)$. If a wavelength filter is located at any point on the system there will be a selection of points at the image.
Viewing the image is equivalent to viewing the pseudoscopic image in a hologram. When the observer moves his head in a one horizontal direction, the parallax change corresponds to moving his head in an opposite direction (in the case of looking directly at the object). When the field of view allows both eyes to observe simultaneously, pseudoscopy by this real-time situation is a very interesting and unique experience. The projected image looks like the pseudoscopic image from a hologram and, in a situation where the object may be in movement, an animated scene is possible. Also, no speckle effects are noticeable, and it is possible to hide the object, making the viewing more impressive to the observer.

**5. Experimental results with multiple wavelengths**

We employed two plastic embossed holographic transmission gratings of the same type, commercially available for architectural or educational purposes, with 533±5 lines/mm sandwiched between two glass plates 2 mm thick. Their effective area employed was less than 60 mm (H) x 40 mm (V). Undulations were evident on both, which prevented us from using their second diffraction order, where light beams appeared distorted. They were located 600±2 mm apart in parallel position and a vertical black paper slit 0.7±0.15 mm wide was in between both gratings. Parallelism of the gratings planes was verified to better than ±1 mm by making coincident reflections of a diode laser beam which traversed the slit, impinged on both gratings and returned to the laser exit. Photographs were made by a analog camera SONY video 8 Handy cam camera connected to a INTEL CS430 web camera whose only purpose was to act as a capture digital converter. It was connected to a Pentium I computer to get 240 x 320-pixel resolution.

For the first object we used a set of three small filament lamps, of the kind employed for illuminating car panels. The filaments were 2 mm long, facing towards the grating. They were arranged in such a way that two of them were at 16 cm from the first grating and their vertical distance was 3 cm, while the third one was 1.5 cm closer to the grating. Fig.4 shows the situation for three viewpoints when the camera moved equal distances from left to right. The light of the image passed from greater to smaller diffraction angle, and consequently the wavelengths, from longer to shorter.

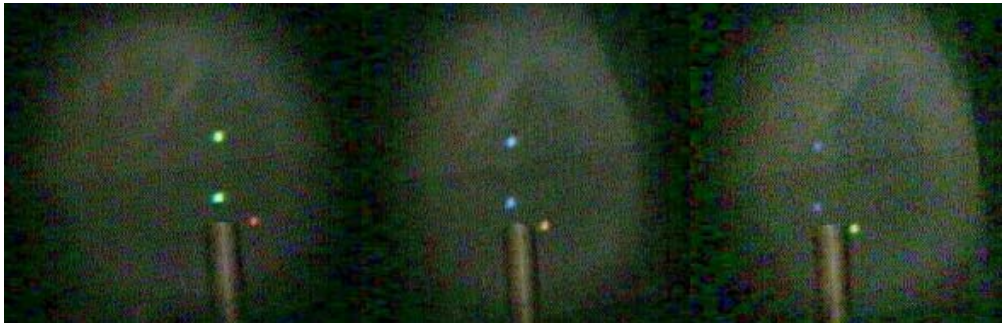

Fig. 4. (2.5 Mb) Parallax and color change for three point objects (see video).
a) left view b) center view c) right view

A rod included in the scene in contact with the second grating serves as a reference for positions. In a) the two vertically aligned lamps appear in green, and the closer lamp in red. In b), the image color of those two lamps is blue, and the image of the other is red. In c) the formerly blue images became in deep blue, and the other image appears now in green. The three spots were slightly displaced from right to left, while the horizontal distance between the spots increased 8 %, a result that shows the depth inversion between image and object in good agreement with the situation (Fig.3).

For the second object, we used a halogeneous 50 W lamp with a parabolic 46 mm diameter faceted reflector behind it, constituting an extended object. The image (see Fig.5) is compared to the image of the object itself, as viewed from the same distance which the light from the object traversed to form the image.

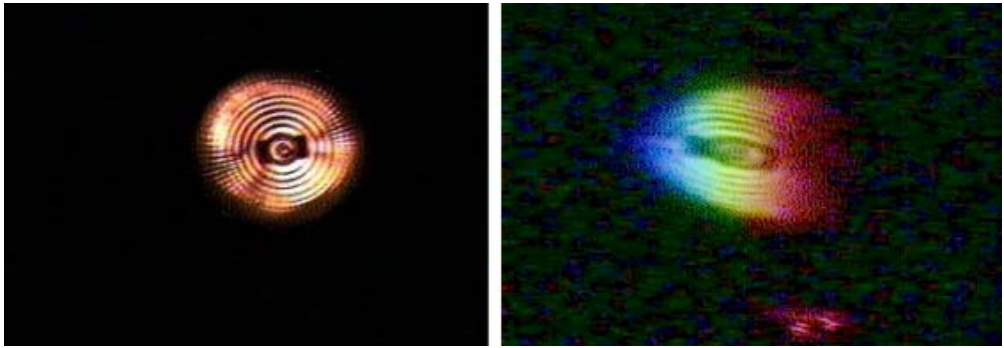

Fig. 5. Image of a halogeneous lamp with reflector. Left: direct image; Right: double diffracted image

The horizontal angular extension of the image was close to that of the object, but did not allow seeing the whole object. The red dot which appears below the image comes from the aligning laser indicating the point at the second grating with zero value for the $x$ coordinate.

We can see, by comparing figures 5 (left) and 5 (right), that the elliptical horizontal extension of the image indicates horizontal magnification of about x1.9.

### 6. Conclusions

We demonstrated a new way of generating a pseudoscopic image directly from an object, which does not needs refracting elements. Also, that image enlargement is possible in one direction by means of purely diffractive element and using the whole spectrum of white light. An aperture which gives a large field for viewpoints comes from a diffractive element whose construction and manipulation is much easier than that of conventional optical elements. The reproduced light field is very similar to the original object field where no magnification distortions are present, even in a longitudinal direction. We showed that white light 3D imaging through diffractive optics render images with an interesting resemblance to holographic images. It is an interesting possibility for increasing the aperture of an optical system because diffraction gratings can be made to deflect light at very large angles generating large angular aperture values.

### 7. Acknowledgements

The "Coordenação de Aperfeiçoamento de Pessoal de Nível Superior"- CAPES of the Brazilian Ministry of Education is acknowledged for a fellowship for Mrs. Noemí R. Rivera. Marcelo F. Rigon is acknowledged for helping in the video registering and digital photographing performed. Diane Marie Petty and Paul and Silvia Baldi are acknowledged for reviewing the English version.